
\font\tenmib=cmmib10
\font\sevenmib=cmmib10 scaled 800
\font\titolo=cmbx12
\font\titolone=cmbx10 scaled\magstep 2
\font\cs=cmcsc10

\font\crs=cmbx8
\font\ninerm=cmr9
\font\ottorm=cmr8
\textfont5=\tenmib
\scriptfont5=\sevenmib
\scriptscriptfont5=\fivei

\font\msytw=msbm10 scaled\magstep1

\font\indbf=cmbx10 scaled\magstep2
\font\type=cmtt10

%
%
%
%
%
%
%

\global\newcount\numsec\global\newcount\numapp
\global\newcount\numfor\global\newcount\numfig\global\newcount\numsub
\numsec=0\numapp=0\numfig=1
\def\veroparagrafo{\number\numsec}\def\veraformula{\number\numfor}
\def\veraappendice{\number\numapp}\def\verasub{\number\numsub}
\def\verafigura{\number\numfig}

\def\section(#1,#2){\advance\numsec by 1\numfor=1\numsub=1%
\SIA p,#1,{\veroparagrafo} %
\write15{\string\Fp (#1){\secc(#1)}}%
\write16{ sec. #1 ==> \secc(#1)  }%
\hbox to \hsize{\titolo\hfill \number\numsec. #2\hfill%
\expandafter{\alato(sec. #1)}}\*}

\def\appendix(#1,#2){\advance\numapp by 1\numfor=1\numsub=1%
\SIA p,#1,{A\veraappendice} %
\write15{\string\Fp (#1){\secc(#1)}}%
\write16{ app. #1 ==> \secc(#1)  }%
\hbox to \hsize{\titolo\hfill Appendix A\number\numapp. #2\hfill%
\expandafter{\alato(app. #1)}}\*}

\def\senondefinito#1{\expandafter\ifx\csname#1\endcsname\relax}

\def\SIA #1,#2,#3 {\senondefinito{#1#2}%
\expandafter\xdef\csname #1#2\endcsname{#3}\else
\write16{???? ma #1#2 e' gia' stato definito !!!!} \fi}

\def \Fe(#1)#2{\SIA fe,#1,#2 }
\def \Fp(#1)#2{\SIA fp,#1,#2 }
\def \Fg(#1)#2{\SIA fg,#1,#2 }

\def\etichetta(#1){(\veroparagrafo.\veraformula)%
\SIA e,#1,(\veroparagrafo.\veraformula) %
\global\advance\numfor by 1%
\write15{\string\Fe (#1){\equ(#1)}}%
\write16{ EQ #1 ==> \equ(#1)  }}

\def\etichettaa(#1){(A\veraappendice.\veraformula)%
\SIA e,#1,(A\veraappendice.\veraformula) %
\global\advance\numfor by 1%
\write15{\string\Fe (#1){\equ(#1)}}%
\write16{ EQ #1 ==> \equ(#1) }}

\def\getichetta(#1){Fig. \verafigura%
\SIA g,#1,{\verafigura} %
\global\advance\numfig by 1%
\write15{\string\Fg (#1){\graf(#1)}}%
\write16{ Fig. #1 ==> \graf(#1) }}

\def\etichettap(#1){\veroparagrafo.\verasub%
\SIA p,#1,{\veroparagrafo.\verasub} %
\global\advance\numsub by 1%
\write15{\string\Fp (#1){\secc(#1)}}%
\write16{ par #1 ==> \secc(#1)  }}

\def\etichettapa(#1){A\veraappendice.\verasub%
\SIA p,#1,{A\veraappendice.\verasub} %
\global\advance\numsub by 1%
\write15{\string\Fp (#1){\secc(#1)}}%
\write16{ par #1 ==> \secc(#1)  }}

\def\Eq(#1){\eqno{\etichetta(#1)\alato(#1)}}
\def\eq(#1){\etichetta(#1)\alato(#1)}
\def\Eqa(#1){\eqno{\etichettaa(#1)\alato(#1)}}
\def\eqa(#1){\etichettaa(#1)\alato(#1)}
\def\eqg(#1){\getichetta(#1)\alato(fig. #1)}
\def\sub(#1){\0\palato(p. #1){\bf \etichettap(#1).}}
\def\asub(#1){\0\palato(p. #1){\bf \etichettapa(#1).}}

\def\equv(#1){\senondefinito{fe#1}$\clubsuit$#1%
\write16{eq. #1 non e' (ancora) definita}%
\else\csname fe#1\endcsname\fi}
\def\grafv(#1){\senondefinito{fg#1}$\clubsuit$#1%
\write16{fig. #1 non e' (ancora) definito}%
\else\csname fg#1\endcsname\fi}
\def\secv(#1){\senondefinito{fp#1}$\clubsuit$#1%
\write16{par. #1 non e' (ancora) definito}%
\else\csname fp#1\endcsname\fi}

\def\equ(#1){\senondefinito{e#1}\equv(#1)\else\csname e#1\endcsname\fi}
\def\graf(#1){\senondefinito{g#1}\grafv(#1)\else\csname g#1\endcsname\fi}
\def\secc(#1){\senondefinito{p#1}\secv(#1)\else\csname p#1\endcsname\fi}
\def\sec(#1){{\S\secc(#1)}}

\def\BOZZA{\bz=1
\def\alato(##1){\rlap{\kern-\hsize\kern-1.2truecm{$\scriptstyle##1$}}}
\def\palato(##1){\rlap{\kern-1.2truecm{$\scriptstyle##1$}}}
}

\def\alato(#1){}
\def\galato(#1){}
\def\palato(#1){}


{\count255=\time\divide\count255 by 60 \xdef\hourmin{\number\count255}
        \multiply\count255 by-60\advance\count255 by\time
   \xdef\hourmin{\hourmin:\ifnum\count255<10 0\fi\the\count255}}

\def\oramin{\hourmin }

\def\data{\number\day/\ifcase\month\or gennaio \or febbraio \or marzo \or
aprile \or maggio \or giugno \or luglio \or agosto \or settembre
\or ottobre \or novembre \or dicembre \fi/\number\year;\ \oramin}
\setbox200\hbox{$\scriptscriptstyle \data $}

%
%
%
\def\ins#1#2#3{\vbox to0pt{\kern-#2 \hbox{\kern#1 #3}\vss}\nointerlineskip}
%
%
%
\newdimen\xshift \newdimen\xwidth \newdimen\yshift

\def\insertplot#1#2#3#4#5{\par%
\xwidth=#1 \xshift=\hsize \advance\xshift by-\xwidth \divide\xshift by 2%
\yshift=#2 \divide\yshift by 2%
\line{\hskip\xshift \vbox to #2{\vfil%
#3 \includegraphics{#4.ps}}\hfill \raise\yshift\hbox{#5}}}


\let\a=\alpha   \let\g=\gamma     \let\d=\delta  \let\e=\varepsilon
  \let\h=\eta       \let\l=\lambda
\let\m=\mu    \let\n=\nu            \let\p=\pi      \let\r=\rho
\let\s=\sigma         
   \let\o=\omega 
      \let\L=\Lambda

\def\\{\hfill\break\hbox{}} \let\==\equiv

\let\io=\infty 

\let\0=\noindent \def\pagina{{\vfill\eject}}

\def\ie{\hbox{\it i.e.\ }}

\def\tende#1{\,\vtop{\ialign{##\crcr\rightarrowfill\crcr
 \noalign{\kern-1pt\nointerlineskip}
 \hskip3.pt${\scriptstyle #1}$\hskip3.pt\crcr}}\,}
\def\otto{\,{\kern-1.truept\leftarrow\kern-5.truept\to\kern-1.truept}\,}

\def\HHH{{\cal H}}

\def\AA{{\cal A}}

\def\T#1{{#1_{\kern-3pt\lower7pt\hbox{$\widetilde{}$}}\kern3pt}}
\def\VVV#1{{\underline #1}_{\kern-3pt
\lower7pt\hbox{$\widetilde{}$}}\kern3pt\,}
\def\W#1{#1_{\kern-3pt\lower7.5pt\hbox{$\widetilde{}$}}\kern2pt\,}

\def\indica{\leaders \hbox to 0.5cm{\hss.\hss}\hfill}
\def\guida{\leaders\hbox to 1em{\hss.\hss}\hfill}

\def\V#1{{\underline #1}}
\def\aa{{\V \a}}\def\hh{{\V h}}\def\HH{{\V H}}\def\AA{{\V A}}
\def\pps{{\V \psi}}\def\oo{{\V \o}}\def\nn{{\V \n}}

\def\qed{\raise1pt\hbox{\vrule height5pt width5pt depth0pt}}

\def\indic{\hbox{\raise-2pt \hbox{\indbf 1}}}

\def\RRR{\hbox{\msytw R}}

 \def\ZZZ{\hbox{\msytw Z}}
 
\def\TTT{\hbox{\msytw T}}

\newcount\mgnf  
\mgnf=0 

\ifnum\mgnf=0
\def\openone{\leavevmode\hbox{\ninerm 1\kern-3.3pt\tenrm1}}%
\def\*{\vglue0.3truecm}\fi
\ifnum\mgnf=1
\def\openone{\leavevmode\hbox{\ninerm 1\kern-3.63pt\tenrm1}}%
\def\*{\vglue0.5truecm}\fi

\newcount\tipobib\newcount\bz\bz=0\newcount\aux\aux=1
\newdimen\bibskip\newdimen\maxit\maxit=0pt


\tipobib=0
\def\9#1{\ifnum\aux=1#1\else\relax\fi}

\newwrite\bib
\immediate\openout\bib=\jobname.bib
\global\newcount\bibex
\bibex=0
\def\verabib{\number\bibex}

\ifnum\tipobib=0
\def\cita#1{\expandafter\ifx\csname c#1\endcsname\relax
\hbox{$\clubsuit$}#1\write16{Manca #1 !}%
\else\csname c#1\endcsname\fi}
\def\rife#1#2#3{\immediate\write\bib{\string\raf{#2}{#3}{#1}}
\immediate\write15{\string\C(#1){[#2]}}
\setbox199=\hbox{#2}\ifnum\maxit < \wd199 \maxit=\wd199\fi}
\else
\def\cita#1{%
\expandafter\ifx\csname d#1\endcsname\relax%
\expandafter\ifx\csname c#1\endcsname\relax%
\hbox{$\clubsuit$}#1\write16{Manca #1 !}%
\else\probib(ref. numero )(#1)%
\csname c#1\endcsname%
\fi\else\csname d#1\endcsname\fi}%
\def\rife#1#2#3{\immediate\write15{\string\Cp(#1){%
\string\immediate\string\write\string\bib{\string\string\string\raf%
{\string\verabib}{#3}{#1}}%
\string\Cn(#1){[\string\verabib]}%
\string\CCc(#1)%
}}}%
\fi

\def\Cn(#1)#2{\expandafter\xdef\csname d#1\endcsname{#2}}
\def\CCc(#1){\csname d#1\endcsname}
\def\probib(#1)(#2){\global\advance\bibex+1%
\9{\immediate\write16{#1\verabib => #2}}%
}

\def\C(#1)#2{\SIA c,#1,{#2}}
\def\Cp(#1)#2{\SIAnx c,#1,{#2}}

\def\SIAnx #1,#2,#3 {\senondefinito{#1#2}%
\expandafter\def\csname#1#2\endcsname{#3}\else%
\write16{???? ma #1,#2 e' gia' stato definito !!!!}\fi}

\bibskip=10truept
\def\hboxto{\hbox to}

\catcode`\{=12\catcode`\}=12
\catcode`\<=1\catcode`\>=2
\immediate\write\bib<
        \string\halign{\string\hboxto \string\maxit%
        {\string #\string\hfill}&%
        \string\vtop{\string\parindent=0pt\string\advance\string\hsize%
        by -1.55truecm%
        \string#\string\vskip \bibskip
        }\string\cr%
>
\catcode`\{=1\catcode`\}=2
\catcode`\<=12\catcode`\>=12

\def\raf#1#2#3{\ifnum \bz=0 [#1]&#2 \cr\else
\llap{${}_{\rm #3}$}[#1]&#2\cr\fi}

\newread\bibin

\catcode`\{=12\catcode`\}=12
\catcode`\<=1\catcode`\>=2
\def\chiudibib<
\catcode`\{=12\catcode`\}=12
\catcode`\<=1\catcode`\>=2
\immediate\write\bib<}>
\catcode`\{=1\catcode`\}=2
\catcode`\<=12\catcode`\>=12
>
\catcode`\{=1\catcode`\}=2
\catcode`\<=12\catcode`\>=12

\def\makebiblio{
\ifnum\tipobib=0
\advance \maxit by 10pt
\else
\maxit=1.truecm
\fi
\chiudibib
\immediate \closeout\bib
\openin\bibin=\jobname.bib
\ifeof\bibin\relax\else
\raggedbottom
\input \jobname.bib
\fi
}

\openin13=#1.aux \ifeof13 \relax \else
\input #1.aux \closein13\fi
\openin14=\jobname.aux \ifeof14 \relax \else
\input \jobname.aux \closein14 \fi
\immediate\openout15=\jobname.aux

\def\biblio{\*\*\centerline{\titolo References}\*\nobreak\makebiblio}

\def\annota#1#2{\footnote{${}^#1$}{{\parindent0pt\baselineskip 0.32truecm
\ottorm#2\vfill}}}

\ifnum\mgnf=0
   \magnification=\magstep0
   \hsize=14.4truecm\vsize=24.truecm
   \parindent=0.3cm\baselineskip=0.45cm\fi
\ifnum\mgnf=1
   \magnification=\magstep1\hoffset=0.truecm
   \hsize=14truecm\vsize=24.truecm
   \baselineskip=18truept plus0.1pt minus0.1pt \parindent=0.9truecm
   \lineskip=0.5truecm\lineskiplimit=0.1pt      \parskip=0.1pt plus1pt\fi

\nopagenumbers
\null\vskip6truecm
\centerline{\titolone On a conjecture for the critical
behaviour of KAM tori}
\vskip1.truecm \centerline{{\titolo
Federico Bonetto$^{*}$ and Guido Gentile$^{\dagger}$}}
\vskip.4truecm
\centerline{{}$^{*}$ Mathematics Department,
Rutgers University, New Brunswick, 08903 NJ}
\vskip.1truecm
\centerline{{}$^{\dagger}$ Dipartimento di Matematica,
Universit\`a di Roma Tre, Roma, I-00146}
\vskip3.truecm

\line{\vtop{
\line{\hskip1.5truecm\vbox{\advance \hsize by -3.1 truecm
\0{\cs Abstract.}
{\it At the light of recent results in literature we review
a conjecture formulated in} Math. Phys. Electron. J.
{\bf 1} (1995), paper 5, 1--13, {\it about the mechanism
of breakdown of invariant sets in KAM problems and
the identification of the dominant terms in the perturbative
expansion of the conjugating function. We show that some
arguments developed therein can be carried out further only
in some particular directions, so limiting a possible future
research program, and that the mechanism of break
down of invariant tori has to be more complicated
than as conjectured in the quoted paper.}} \hfill} }}

\pagina
\pageno=1
\footline={\hss\tenrm\folio\hss}
\section(1,Introduction)

\sub(1.1) {\it The state of the art of \cita{GGM}.}
In \cita{GGM} a conjecture about the mechanism of breakdown of KAM
invariant tori is proposed. Roughly it is based on the following idea
(we refer to \cita{GGM} for a more detailed and technical exposition
and also for the introduction of the notions
used in the following analysis).

If $\TTT=\RRR/2\p\ZZZ$, let $\TTT^{\ell}$ be the
$\ell$-dimensional torus. Consider a Hamiltonian system
$$ \HHH = \oo_0\cdot\AA + {1\over 2}\AA\cdot J^{-1}\AA
+ \e f(\aa) \; , \Eq(1.1) $$
where $(\aa,\AA)\in \TTT^{\ell}\times\RRR^{\ell}$ are conjugate
variables, $J$ is the matrix of momenta of inertia,
$\cdot$ denotes the inner product in $\RRR^{\ell}$,
$f(\aa)$ is a trigonometric polynomial in the
angle variables and $\e$ is a parameter.

For concreteness one can suppose that $\HHH$ describes
the Escande-Doveil pendulum, \cita{ED}: $\ell=2$ and
$f(\aa)=a\cos\a_1+b\cos(\a_1-\a_2)$, with $(a,b)\in\RRR^2$.

The solutions of the equations of motion
describing the invariant tori for the system \equ(1.1)
with Diophantine rotation vector $\oo$ can be parameterized as
$$ \aa = \pps + \hh (\pps;\e) \; , \qquad
\AA = \AA_0 + \HH(\pps;\e) \;, \Eq(1.2) $$
where $\pps\in \TTT^{\ell}$, $\AA_0=J(\oo-\oo_0)$ and $\hh,\HH$,
for $\e$ small enough, are analytic functions,
whose series expansions admit a graph representation in terms of trees;
we refer to \cita{GGM} for details and definitions.

One can consider also trees in which no resonances (see \cita{GGM},
\S 4) are allowed
to appear but the perturbative parameter $\e$ is replaced by
$\h_\e\=\e(1-\s_\e)^{-1}$, with a suitable (matrix) {\it form factor}
$\s_{\e}$ taking into account the resummation of all resonances,
\cita{GM}, and denote by $\hh^*,\HH^*$ the functions so obtained;
the series expansion in $\h$ will be called {\it resummed series}.

If $\r$ is the radius of convergence of the series defining
$\hh,\HH$ in \equ(1.2) and $\e_c\in\RRR^+$ is the (positive)
critical value at which the tori break down, one has $\e_c\ge \r$;
in general the analyticity domain in $\e$ of the series
defining $\hh,\HH$ is not a circle and it can happen that $\e_c>\r$.

One can imagine that the following scenario arises: the singular
behaviour of $\hh(\pps;\e),\HH(\pps;\e)$ as functions of $\e$ is the
same of that of $\hh^*(\pps;\h_\e),\HH^*(\pps;\h_\e)$ and while, for
$\e\to\e_c^{-}$, $\s_\e$ is still finite, $\h=\h_\e$ goes out the
convergence domain (in $\h$) of the series for $\hh^*,\HH^*$.  If
moreover the analyticity domain in $\h$ of $\hh^*,\HH^*$ turns out to
be a circle this would mean that $\h$ is a more natural parameter for
the invariant tori.

If this really happens then the behaviour of the series near the
critical value is determined by the trees without resonances, for which
the perturbative series would still be meaningful for $\e$ near $\e_c$.

Under such an assumption a universal behaviour of the series
near the critical value is proposed in \cita{GGM}.
Predictions of the value of the critical exponente $\d$ are made,
conjecturing that only some simple classes of trees are relevant:
the linear trees presenting the largest
possible number of small divisors (see \cita{GGM}, \S 5).

Considering (for concreteness purposes) a rotation vector $\oo=(r,1)$,
where $r$ is the golden mean (see \sec(3.1) below),
then one was led to expect that, by denoting by $\{q_n\}$
the denominators of the convergents defined by the
continuous fraction expansion for $r$, a suitable function
$Z(\Lambda_{q_n})$, defined in equation (5.2) of \cita{GGM}
and recalled in \sec(3.1) below, satisfied the asymptotics
$$ \left| Z(\Lambda_{q_n}) \right| \approx
{C^{q_n} \over q_n^{\delta}} \; , \qquad C=C(\h,f) \; , \Eq(1.3) $$
for some positive constant $C(\h,f)$ depending on $\h f$.

\*

\sub(1.2) {\it About the standard and semistandard maps.}
If true, the above mechanism should work also for the standard map,
which is the dynamical system generated by the iteration of the
area-preserving map of the cylinder to itself
$$ \cases{
x' = x + y + \e \sin x \; , & \cr
y' = y + \e \sin x \; , & \cr} \Eq(1.4) $$
where $(x,y)\in \TTT\times\RRR$.
At least it is generally assumed that area-preserving maps and
Hamiltonian flows share the same critical behaviour:
in particular they are expected to have the same critical exponent
$\delta$; see also \cita{GGM}, \cita{CGJ} 
and comments therein about \cita{M1}, \cita{M2}.

The r\^ole of the functions $\hh,\HH$ is now played by
two scalar functions $u,v$ (see, for instance, \cita{BG1}) such that
$$ x=\a+u(\a,\e) \; , \qquad y = 2\p\o+v(\a,\e) \; , \Eq(1.5) $$
with $\o$ a fixed rotation number and $u,v$
analytic in their arguments
\annota{1}{The functions $\scriptstyle u,v$ are trivially related:
$\scriptstyle v(\a,\e)=u(\a,\e)-u(\a-2\p\o,\e)$.}.
In terms of $\a\in \TTT$, the
dynamics is a trivial rotation: $\a\to\a'=\a+2\p\o$.

Analogously to the Hamiltonian case \equ(1.1),
one can define two functions $u^*,v^*$,
obtained by considering only the values of the trees contributing
to $u,v$ without resonances (and replacing $\e$ with $\h$, where
$\h\=\h_\e=\e(1-\s_\e)^{-1}$ takes into account the resummation
of resonances; note that now $\s_\e$ is a scalar).

For the standard map, let us denote by $\r(\o)$ and $\e_c(\o)$,
respectively, the radius of convergence (in $\e$)
and the critical value for fixed rotation number $\o$.

Consider also the semistandard map, introduced by Chirikov, \cita{C},
$$ \cases{
x' = x + y + \e' \exp i x \; , & \cr
y' = y + \e' \exp ix \; , \qquad & $ \e'=\e/2 \; , $ \cr} \Eq(1.6) $$
and call $u_0,v_0$ the corresponding conjugating functions; let us
denote by $\r_0(\o)$ the radius of convergence (in $\e'$)
for the semistandard map\annota{2}{
We could define also the functions $\scriptstyle u_0^*,v_0^*$
and the critical value $\scriptstyle\e_{0c}(\o)$ for the semistandard map,
but one has $\scriptstyle u_0=u_0^*,v_0=v_0^*$, as the semistandard map has
no resonances, and $\scriptstyle\e_{0c}(\o)=\r_0(\o)$:
to deduce the latter property simply note that the semistandard map
is invariant under rotation of $\scriptstyle\e'$ in the complex plane.}.

\*

\sub(1.3) {\it The implications of \cita{D1}.}
Given a rotation number $\o\in(0,1)$, let $B(\o)$ be the function
$$ B(\o) = \sum_{n=0}^{\io} {\log q_{n+1} \over q_n} \; , \Eq(1.7) $$
where $\{q_n\}$ are the convergents defined by the continued fraction
expansion for the rotation number $\o=[a_1,a_2,a_3,\ldots]$,
so that $q_n=a_nq_{n-1}+q_{n-2}$, for $n\ge 1$, and $q_{-1}=0$, $q_0=1$.
The function \equ(1.7) is related to the
Bryuno function introduced by Yoccoz, \cita{Y}.

Consider the series defining $u^*\=u^*(\a,\h)$.
Of course one can write
$$ u^*(\a,\h) = \sum_{k=1}^{\io} \h^k
\left( \sum_{\n=-k}^{k} u^{*(k)}_\n e^{i\n\a} \right) 
\= \sum_{k=1}^{\io} \h^k u^{*(k)}(\a) \; . \Eq(1.8) $$
The radius of convergence $R(\o)$ of a series like \equ(1.8)
can be expressed as
$$ R^{-1}(\o) = \sup_{\a\in[0,2\p]} \limsup_{k\to\io}
\left| u^{*(k)}(\a) \right|^{1/k} =
\limsup_{k\to\io} \max_{|\n|\le k} \left| u^{*(k)}_\n \right|^{1/k}
\; , \Eq(1.9) $$
as the equivalence between the two definitions has been shown in \cita{D1}.

On the other hand the coefficient in \equ(1.8) with $\n=k$
is the same for both the functions $u$ and $u^*$ (and one has
$u^{(k)}_k=u^{*(k)}_k=u^{(k)}_{0k}$).
Moreover it is easy to see that one has
$$ \left| u^{(k)}_{0k} \right| \ge B_1^k e^{2B(\o)k}\Eq(1.10) $$ 
for $k$ large enough and some constant $B_1$; again see \cita{D1}.

This implies that
$$ \limsup_{k\to\io}
\max_{|\n|\le k} \left| u^{*(k)}_\n \right|^{1/k}
\ge B_1 e^{2B(\o)} \; , \Eq(1.11) $$
%
so that $R(\o)\le B_1^{-1}e^{-2B(\o)}$,
\ie the radius of convergence in $\h$ of the function $u^*$
at best should be of order of $\r_0(\o)$
\annota{3}{In the same way one finds that the radius of convergence in 
$\scriptstyle\e$
of the function $\scriptstyle u$ at best should be of order of 
$\scriptstyle\r_0(\o)$.}.

\*

\sub(1.4) {\it The implications of \cita{D2}}.
One has $\r_0(\o)=C_1(\o)e^{-2B(\o)}$ with $C_1(\o)$
satisfying the bound $C_{1}^{-1}<C_1(\o)<C_{1}$, uniformly in $\o$,
for a suitable constant $C_{1}$, \cita{D1}.
Also for the standard map one has the same
dependence of the radius of convergence $\r(\o)$ on the
function $B(\o)$, with a possibly different function
$C_1(\o)$, always admitting a bound from below and from above:
the bound from above is proven in \cita{D1}, by using the
argument recalled in \sec(1.3), and the one from below in 
\cita{BG2}.

In \cita{D2} it is proven that for rotation numbers
\annota{4}{With the notations in \sec(1.1), one has $\scriptstyle r=\g_1$.}
$\o=\g_n\=[n,n,n,\ldots]$, with $n$ large enough,
one has $C_2/n > \e_c(\g_n)>C_2^{-1}/n$, for some constant $C_2$;
as $B(\o)> \log n$, for $\o=\g_n$, then $\e_c(\g_n)>C_2^{-1}n^{-1}$
and $\r(\g_n)<C_1 n^{-2}$, (see \cita{D2}, end of section 5).
Therefore one can choose $n$ so large that $\r(\g_n)/\e_c(\g_n)$
is smaller than any prefixed quantity.

Note that the difference between the radius of convergence
and the critical value becomes relevant only for some rotation numbers
(like the noble numbers $\g_n$, with $n$ large enough).
For instance for $\g_1$, according to numerical simulations,
the analyticity domain appears (very) slightly stretched
along the imaginary axis so that not only the two quantities
are comparable, but even $\e_c(\g_1)=\r(\g_1)$, \cita{FL}.

\*\*
\section(2,Discussion)

\sub(2.1) {\it About the \equ(1.3).} 
If the dominant contributions to $u^*,v^*$ were given by summing
only the values of trees defining the functions $Z(\Lambda_{q_n})$,
as conjectured in \cita{GGM},
then the standard map and the semistandard map should have the
same critical behaviour. In fact in terms of trees the semistandard map
admits the same graph representation of the standard map,
with the only (remarkable) difference that the mode
labels have all the same signs;
then the paths with the maximal number of small
divisors in $Z(\Lambda_{q_n})$ are the same for both the
standard map and the semistandard map.

This means that, accepting the above conjecture, one ought to expect
that the radius of convergence (in $\e$) of the functions $u_0,v_0$
for the semistandard map and the radius of convergence (in $\h$)
of the functions $u^*,v^*$ for the standard map should be
equal to each other and, in particular, of size of $\r_0(\o)$.

Furthermore the results listed in \sec(1.3) imply that the
functions $u^*,v^*$ can not have a radius of convergence
larger than that of the functions $u_0,v_0$,
also considering all trees contributing to them and not only
the ones defining the function $Z(\L_{q_n})$.

In other words the semistandard map should capture all the critical
behaviour of the standard map: if so there would be contradiction
with the results existing in literature. As a matter of fact
we shall see in \sec(3) that the behaviour \equ(1.3)
conjectured for $Z(\L_{q_n})$ does not hold.

\*

\sub(2.2) {\it About the behaviour of $\s_\e$ near $\e_c$.}
The idea that the critical behaviour
can be studied through the series obtained be neglecting
the resonances (that is by resumming them and defining
a new parameter $\h\=\h_\e$, as briefly recalled in \sec(1.1)
and in \sec(1.2)) is presented in \cita{GGM}.
One could also interpret the numerical
results of \cita{CGJ}, \cita{CGJK} as supporting such an idea:
the fact that the KAM iteration represents the good
transformation to look at also far from the KAM
analyticity domain could suggest that a perturbative
approach to the study of the breakdown of KAM invariant tori
in some sense should be possible: this was the idea underlying
the analysis performed in [GGM]. Anyway much work has
still to be done in this direction; see also \sec(4) below.

 From the results listed in \sec(1) the following picture emerges.
For $\e$ small enough $|\s_{\e}|< R|\e|$ for some constant $R$;
see the theorem in \cita{GGM}, \S 4. When $\e$ grows
along the real axis toward the critical value, the
possibility to have $\s_\e$ still finite for $\e=\e_c$
is consistent with a perturbative approach
if one of the following two cases arise:\\
(1) if $\s_\e$ is finite and smooth in $\e$ for $\e=\e_c$, then it
can happen that $|\h_c|\=|\e_c(1-\s_{\e_c})^{-1}|$ becomes equal
to the radius of convergence $R(\o)$ of the functions $u^*,v^*$;\\
(2) if $\s_\e$ is finite and singular in $\e$ for $\e=\e_c$,
while $\h_c=\e_c(1-\s_{\e_c})^{-1}$ is smaller than the
radius of convergence $R(\o)$ of the functions $u^*,v^*$, then
the singularity of the tori shows up through the
singular dependence of $\s_\e$ on $\e$ at the critical value.\\

In both cases the perturbative series for $u^*,v^*$
can be used also near the critical value, because for
any $|\e|<\e_c$ one has that $\h$ is smaller than
the radius of convergence of the series.

In the case of $\o=\g_n$, in principle two possibilities can be
envisaged for the behaviour of $\s_{\e}$ for $\e$ near $\e_c$:\\
(i) either $|\s_\e|<1$,\\
(ii) or else $\s_\e$ negative and $|\s_\e|\ge 1$
\annota{5}{Note that $\scriptstyle\s_\e$ positive and 
$\scriptstyle|\s_\e|\ge 1$ is not possible as
the analyticity of $\scriptstyle\s_\e$ in $\scriptstyle\e$ for small 
$\scriptstyle\e$ would imply the existence of a value
$\scriptstyle\bar\e$ such that $\scriptstyle|1-\s_{\bar\e}|=0$.}.\\

Consider $\o=\g_n$, with $n$ large enough, and
assume the results in \cita{D2}.
Then, if the analyticity domain in $\h$ was a circle
with radius $R(\o)$ and, for $\e\to\e_c^-(\o)$, $|\s_\e|<1$
and $\h\to R(\o)$, one should have
$|\h|=|\e(1-\s_\e)^{-1}|\to R(\o) \approx \r_0(\o)$,
hence $|1-\s_\e|\to |1-\s_{\e_c}| \approx \e_c(\o)/\r_0(\o)
>C_0n$, with $C_0^{-1}=C_1C_2$,
which is incompatible with $|\s_\e|<1$.
So only the possibility (ii) above could be consistent
both with [D2] and with the scenario proposed in \cita{GGM},
for $\o=\g_n$.

More generally, for $\o=\g_n$, with $n$ large enough,
if one had $\h\to \h_c\le R(\o)$ for $\e\to\e_c^-(\o)$,
then $|\s_\e|$ should become at least of order
$e^{B(\o)}$ for $\e\to\e_c^-(\o)$, \ie $|\s_\e|\to |\s_{\e_c}|
\ge C_3(\o)e^{B(\o)}$,
for some function $C_3(\o)$ bounded uniformily in $\o$:
in fact in this way one would have $|\h_c|\=
|\e_c(1-\s_{\e_c})^{-1}| \approx C_3\e_c(\o)e^{-B(\o)}\approx
C_4\r_0(\o)=R(\o)$, for some constants $C_3,C_4$,
in case (1) and $|\h_c|< C_4\r_0(\o)=R(\o)$ in case (2).

\*\*
\section(3,Analytic results)

\sub(3.1) {\it The definition of $Z(\L_{q_n})$.}
If $r=(\sqrt{5}-1)/2$ is the {\it golden mean},
let us call $\{p_n/q_n\}$ the convergents of the continuous fraction
expansion for $r$, where $\{p_n\}$ is the {\it Fibonacci sequence}
defined by $p_{n+1}=p_n+p_{n-1}$ with $p_{-1}=1$ and $p_0=0$,
so that $p_n=q_{n+1}$ with $q_{-1}=0$ and $q_0=1$.

The sequence of numbers $Z_n\=Z(\Lambda_{q_n})$ is defined
in \cita{GGM}, equation (5.2), as sum of the values
of a suitable class of trees which can be described by
the family $\L_{q_n}$ of self-avoiding walks on $\ZZZ^2$
starting at $(0,0)$, ending at $(q_n,-p_n)$ and contained in
the strip $0< x< q_n$, except for the left extreme points.

Then the numbers $Z_n$ can be approximately defined by
the recursive relation
$$ Z_{n+1}=Z_{n}Z_{n-1}\left({\e_{n-1}\over \e_{n+1}}\right)^2
\; , \Eq(3.1) $$
where, for consistency, we fix $Z_{-1}=r^{2}$ and 
$Z_{0}=r^{-2}$, \cita{GGM}. Set also $\e_n=q_n r-p_n$;
then $\e_{-1}=-1$ and $\e_n=q_n r-q_{n-1}$.

Instead of studying \equ(3.1) define $\lambda_n=\log Z_n$, so that
$$ \l_{n+1}=\l_{n}+\l_{n-1}+2 \left( \log\e_{n-1}-\log\e_{n+1}
\right) \; , \Eq(3.2) $$
with $\l_{-1}=2\log r$ and $\l_{0}=-2\log r$.

\*

\sub(3.2) {\it Against the \equ(1.3).}
All the following identities can be easily proven by induction using
the fact  that $r$ is the positive solution of $r^2+r-1=0$.

First note that $\e_n=(-1)^nr^{n+1}$. It follows that
$$ \l_{n+1}=\l_{n}+\l_{n-1}-4\log r\; . \Eq(3.3) $$
which satisfies $\l_n=\m_n+ s_n$, where
$$ \cases{
s_{n} = - 4\log r\sum_{i=0}^{n-1} q_{n-1-i} \; , & for $n\ge 1 \; , $ \cr
s_{0} = s_{-1}= 0 \; , & \cr} \Eq(3.4) $$
and $\m_n$ is defined by $\m_{n+1}=\m_{n}+\m_{n-1}$,
with $\m_{-1}=2\log r$ and $\m_{0}=-2\log r$.
It is now immediate that $\m_n=-\left( 2\log r \right) q_{n-2}$, for
$n\ge 1$, so that we have
$$ \l_{n}=-2\log r \left(q_{n-2}+2\sum_{i=0}^{n-1} q_i\right)\; ,
\Eq(3.5) $$
for $n\ge 1$.

Let now use that
$$ q_n={r^{-(n+1)}-(-1)^{n+1}r^{n+1}\over r+r^{-1}}
\; .\Eq(3.6)$$
Inserting this expression in \equ(3.4) we get, for $n\ge 1$,
$$ \eqalign{\l_n&={2\log r^{-1}\over r+r^{-1}}
\left(r^{-(n-1)}+2\sum_{i=0}^{n}r^{-i} -
(-1)^{n-1}r^{n-1} - 2\sum_{i=0}^{n}(-1)^ir^i\right) = \cr
&=2\log r^{-1}\left(r^{-(n+2)}-2+(-1)^{n+2}r^{n+2}\right)
\; , \cr}\Eq(3.7) $$
which can be written as
$$\l_n = 2\log r^{-1}\left[(r+r^{-1})q_{n+1}-2+2(-1)^{n+2}r^{n+2}
\right] \; ,\Eq(3.8) $$
which means
$$Z_n=r^{4}r^{-2(r+r^{-1})q_{n+1}}r^{-4(-1)^{n+2}r^{n+2}} \; . \Eq(3.9) $$
This implies, using that
$$ r^{n+1}={ 1 \over r+r^{-1} }
q_n^{-1} \left[ 1 + (-1)^{n} r^{2(n+1)} \right] \; , \Eq(3.10) $$
the following asympotics:
$$ \left| Z(\Lambda_{q_n}) \right| \approx KC^{q_{n+1}}\left(1+ 
(-1)^ndq_{n+1}^{-1}\right) \; , \Eq(3.11) $$
with $K=r^{-4}$, $C=r^{-2(r+r^{-1})}$ and 
$d=4(r+r^{-1})^{-1}\log r^{-1}$.
This is clearly in contradiction with \equ(1.3).

\*\*
\section(4,Conclusions)

\sub(4.1) {\it Dominant contributions.}
Our attention to \cita{GGM} has been called back by the
recent papers \cita{CGJ}, \cita{CGJK},
where the breakdown of KAM invariant tori
(for two-dimensional Hamiltonian systems)
is numerically studied through a renormalization group scheme,
and by the results in \cita{D2}.


Even if it is not true that the terms
defining $Z(\Lambda_{q_n})$ are the only relevant ones
(as the results in \sec(3) show),
one can argue that the real dominant contributions
are given by the trees having the mode labels accumulating
near the resonant line (\ie such that the small divisors
are really as small as possible),
but not necessarily belonging to the class described by $\L_{q_n}$.

In fact let us compare the tree values for the Escande-Doveil pendulum
with the ones for the standard map and accept that they have the
same singular behaviour at the critical value (as it is
generally believed, \cita{ED}): the small divisors are
respectively $(i\oo\cdot\nn)^{2}$, $\nn=(\n_1,\n_2)\in\ZZZ^2$,
and $2[\cos(2\p\o\n)-1]$, $\n\in\ZZZ$. So one can note that
they have the same smallness problem, with the only
difference that for the Escande-Doveil pendulum
one can have also small divisors which are not small at all
(when $\nn$ is nearly parallel to $\oo$), while for the
standard map one can approximate the quantity
$2[\cos(2\p\o\n)-1]$ with something of the form
$(i(\o\n_1+\n_2))^2$, with $\n_1=\n$ and
$\n_2$ such that $|\o\n_1+\n_2|\le 1$:
in other words one can expect that the tree expansion
for the standard map is very similar to that of the
Escande-Doveil pendulum, but it does contain only
the trees with the momenta directed along the resonant line.
Of course this imply neither that in the case of linear trees
the most dominant contributions are the ones with
the mode labels having all the same signs
nor that only the linear trees are relevant.
As a matter of fact the analysis performed in \sec(3)
shows that making a so restrictive assumption leads to wrong results.

\*

\sub(4.2) {\it Resummed series: smooth form factors.}
In conclusion an overall behaviour like \equ(1.3)
could be still possible in principle, even if a larger
class of trees ought to be taken into account.

Let us consider $\o=\g_n$, with $n$ large enough
and use the results in \sec(2.2).
If for $\e\to\e_c$ one had $|\s_\e|<1$, then we have seen
that the analyticity domain in $\h$ of $u^*,v^*$
can not be a circle: rather, defining
$\h_c(\o)$ as the equivalent of $\e_c$ for the functions
$u^*,v^*$, the ratio $\h_c(\o)/R(\o)$ would be of the
same order of the ratio $\e_c(\o)/\r(\o)$, and
no simplification could arise in considering the
resummed series $u^*,v^*$ instead of the original ones $u,v$.
In other words $\h$ would not be a ``natural'' parameter.

Of course in the case of the standard map,
deep cancellation mechanisms should intervene in order to enlarge
the analyticity domain along the real axis
of the series for $u^*,v^*$ from
$R(\o)\approx \r_0(\o)$ to a quantity $\h_c(\o)$
of the same order of $\e_c(\o)$.
Analogous considerations could be made for the functions
$\hh^*,\HH^*$ in the case of Hamiltonian flows.

On the contrary if, for $\e\to\e_c$,
one had case (1), possibilty (ii) -- see \sec(2.2) --,
so that $\s_\e\to C_2(\o)e^{B(\o)}$,
the convergence domain in $\h$ for $u^*,v^*$ could still be a circle.
Of course, if this is the case, the study of the factor form $\s_\e$
could turn out to be a very difficult task:
in fact it could become a nonperturbative problem, as the full
dependence on $\e$ would be required for $\e\to\e_c$.

A perturbative analysis could still be possible
for some rotation number, for instance for $\o=\g_1$, when
the radius of convergence and the critical
value are expected to be equal (see comments at the
end of \sec(1.4)); anyway we have no evidence of this,
so we leave it as an open problem.

\*

\sub(4.3) {\it Resummed series: singular form factors.}
Suppose now that at the critical value $\s_\e$ is finite and singular.
One can consider the resummed series for the form factor
$\s_\e$, by expressing it as series in $\h$,
\ie $\s_\e=F(\h)=F(\e(1-\s_\e)^{-1})$, where the function $F$
admits a graph representation in terms of trees without resonances
and with $\e$ replaced with $\h_\e$.
If for $\e=\e_c$ one has that $\h$ is inside (enough) the domain
in which the perturbative series for the function $F$ converges,
then one can try to truncate the series to the first orders,
so obtaining an (approximate) implicit equation for $\s\=\s_\e$:
the solution should be singular at the value $\e=\e_c$.
So the possibility of a perturbative study of the breakdown
of KAM invariant curves has not to be excluded
\annota{6}{Note that for $\scriptstyle\o=\g_n$, with $\scriptstyle n$ 
large enough, one can still
define the function $\scriptstyle F(\h)$: the problem is that for 
$\scriptstyle\e$ near to the
critical value no truncation of the series would be meaningful.}.

Further investigation (also from a numerical point of view)
in this direction would be highly profitable. Also
a comparison between the values of the involved quantities
for the standard map and the semistandard map would
be very enlightening.

\*\*

\0{\bf Acknowledgments.}
We want to thank the ESI in Vienna, where part of this work
has been done, for hospitality. We also thank V. Mastropietro
for interesting discussions and in particular G. Gallavotti
for continuous enlightening suggestions and critical comments. 

\*\*
\appendix(A1,Numerical analysis)

\asub(A1.1) {\it Motivations.}
Although the results of \sec(3) are conclusives we report here on some
numerical simulations that suggested us those results. 
We tried to fit $\l_n$ with a slightly more general
relation than \equ(1.3), \ie $\l_{n}=k+\d\log q_n +cq_n$.
The problem of this fit is that $q_n \simeq r^{-n}$ ,so that we have
to compute  all the involved quantities with a very high accuracy:
if not the constant $k$ and the ``linear'' term $\d$ will be completely lost.
We think that the way we did can be of some interest for the reader.

We used the standard Unix command $\type bc$ that is able to execute
computation written in a C-like programming language that operates
in fixed point notation with number of arbitrary dimension
and an arbitrary but prefixed precision. We compute
$c_i,k_i,\d_i$ has the best square fit of \equ(3.1) using the value of
$\l_m$ with $m=(i-1)100$ to $m=i100$. The only problem is to choose
the precision at which the computation are done in such a way to have
a desired precision in the value of $c_i,k_i,\d_i$. This is what we
will discuss in the next subsection.

\*

\asub(A1.2) {\it About the precision.}
Let $p$ be our chosen precision, \ie we make all operation with
$p$ significant digits after the point. This mean that at every elementary 
operation creates an error $O(10^{-p})$.

%
%
%
%

If we call $\theta_i$ the error due to this round off at step $i$ 
it's easy to see that the accumulated error on $\l_n$ is 
$\sum_{i=0}^{n-1}\theta'_iq_{n-1-i}=O(10^{-p}r^n)$.
To compute the best quadratic fit we have to
solve the equation ${\bf A}\vec\chi=\vec v$, where ${\bf A}$
is the symmetric matrix ${\bf A}=\sum_{i=0}^{n}\vec Q_n\otimes
\vec Q_n$, with $\vec Q_n=(q_n,\log(q_n),1)$,
$\vec v=\sum_{i=0}^{n}\vec Q_n\l_n$ and $\vec
\chi=(c,\delta,k)$.  This mean that ${\bf A}_{1,1}\approx r^{-2n}$,
while ${\bf A}_{1,2},{\bf A}_{1,3}\approx r^{-n}$ and all the other
entries ${\bf A}_{i,j}$ of ${\bf A}$, with $i\ge j$, are of order 1.

Observing that $({\bf A}^{-1})_{i,j}={\bf A}^{i,j}/\det{\bf A}$ and 
that $\det {\bf A}\approx r^{-2n}\pm 10^{-p}r^{-2n}$, we get
$$\left\{
\eqalign{
&({\bf A}^{-1})_{1,1}\approx r^{2n}\pm 10^{-p} \; , \cr
&({\bf A}^{-1})_{1,3},({\bf A}^{-1})_{1,2}\approx 
r^{n}\pm 10^{-p}r^{-n} \; , \cr
&({\bf A}^{-1})_{2,2}\;,\;({\bf A}^{-1})_{2,3}\;,\;
({\bf A}^{-1})_{3,3} \approx 1 
\pm 10^{-p}r^{-2n} \; . \cr}  \right. \Eq(4.4) $$ 

The above estimates with the fact that $\vec v\approx
(r^{-2n}\pm 10^{-p}r^{-2n},r^{-n}\pm 10^{-p}r^{-n},
r^{-n}\pm 10^{-p}r^{-n})$ implies an error on $\chi$
of order $10^{-p}r^{-2n}$.

\*

\asub(A1.3) {\it Numerical results.}
The numerical result that we get are summarized in the next
table, which, in our opinion, requires no comments.
According to our error analysis, we fixed the precision to
$p=2000\log_{10}\omega+20$ in such a way that the reported numbers
should be reliable within 20 digits. The agreement with the 
analytical evaluation reported in \sec(2) is perfect.

$$\vcenter{\halign{\strut\vrule\kern 3mm\hfill#\hfill\kern 3mm&
\vrule\kern 3mm\hfill#\hfill\kern 3mm&
\vrule\kern 3mm\hfill#\hfill\kern 3mm&
\vrule\kern 3mm\hfill#\hfill\kern 3mm\vrule\cr
\noalign{\hrule}
$n$&$c$&$\delta$&$k$\cr
\noalign{\hrule}
100&3.482081477708027334&-.000312407255699678&-1.915302823598781109\cr
\noalign{\hrule}
200&3.482081477708027334&-.000000000000000000&-1.924847300238413790\cr
\noalign{\hrule}
300&3.482081477708027334&-.000000000000000000&-1.924847300238413790\cr
\noalign{\hrule}
400&3.482081477708027334&-.000000000000000000&-1.924847300238413790\cr
\noalign{\hrule}
500&3.482081477708027334&-.000000000000000000&-1.924847300238413790\cr
\noalign{\hrule}
600&3.482081477708027334&-.000000000000000000&-1.924847300238413790\cr
\noalign{\hrule}
700&3.482081477708027334&-.000000000000000000&-1.924847300238413790\cr
\noalign{\hrule}
800&3.482081477708027334& .000000000000000000&-1.924847300238413790\cr
\noalign{\hrule}
900&3.482081477708027334& .000000000000000000&-1.924847300238413790\cr
\noalign{\hrule}
1000&3.482081477708027334& .000000000000000000&-1.924847300238413790\cr
\noalign{\hrule}
}}$$


\line{\hfill\vbox{\advance \hsize by -3.1 truecm
\0{\baselineskip=0.30truecm\crs Table 1. Numerical results.
\ottorm The small deviations for $\scriptstyle{i=1}$ is due
to the fact that our fit for $\scriptstyle{\l_n}$ contains 
$\scriptstyle{q_n}$ instead of $\scriptstyle{q_{n+1}}$.\vfill}} \hfill}

\rife{BG1}{BG1}{A. Berretti, G. Gentile:
Scaling properties for the radius of convergence of a
Lindstedt series: the standard map,
{\it J. Math. Pures Appl.} {\bf 78} (1999), 159--176. }
\rife{BG2}{BG2}{A. Berretti, G. Gentile:
Bryuno function and the standard map, preprint. }
\rife{C}{C}{V. Chirikov:
A universal instability of many-dimensional oscillator systems,
{\it Phys. Rep.} {\bf 52} (1979), 263--379. }
\rife{CGJ}{CGJ}{C. Chandre, M. Govin, H.R. Jauslin:
Kolmogorov-Arnold-Moser renormalization-group approach to
the breakup of invariant tori in Hamiltonian systems,
{\it Phys. Rev. E} {\bf 57} (1998), (2), 1536--1543. }
\rife{CGJK}{CGJK}{C. Chandre, M. Govin, H.R. Jauslin, H. Koch:
Universality of the breakup of invariant tori in Hamiltonian flows,
{\it Phys. Rev. E} {\bf 57} (1998), (6), 6612--6617. }
\rife{D1}{D1}{A.M. Davie:
The critical function for the semistandard map,
{\it Nonlinearity} {\bf  7} (1994), 219--229. }
\rife{D2}{D2}{A.M. Davie:
Renormalization for analytic area-preserving maps, unpublished. }
\rife{ED}{ED}{D.F. Escande, F. Doveil:
Renormalization method for computing the threshold of the large-scale
stochastic instability in two degrees of freedom Hamiltonian systems,
{\it J. Stat. Phys.} {\bf 26} (1981), 257--284. }
\rife{FL}{FL}{C. Falcolini, R. de la Llave:
Numerical calculation of domains of analyticity for
perturbation theories with the presence of small divisors,
{\it J. Stat. Phys.} {\bf 67} (1992), 645--666. }
\rife{GGM}{GGM}{G. Gallavotti, G. Gentile, V. Mastropietro:
Field Theory and KAM tori,
{\it Math. Phys. Electron. J.} {\bf 1} (1995), paper 5, 1--13. }
\rife{GM}{GM}{G. Gentile, V. Mastropietro:
Tree expansion and multiscale decomposition for KAM tori,
{\it Nonlinearity} {\bf 8} (1995), 1--20. }
\rife{M1}{M1}{R.S. MacKay:
A renormalization approach to invariant circles in area-preserving maps,
{\it Physica D} {\bf  7} (1983), 283--300. }
\rife{M2}{M2}{R.S. MacKay:
{\sl Renormalization in area-preserving maps},
World Scientific, London (1993). }
\rife{Y}{Y}{J.-C. Yoccoz:
Th\'eoreme de Siegel, nombres de Brjuno and p\^olinomes quadratiques,
{\it Ast\'erisque} {\bf 231}, 3--88 (1995). }

\biblio

\bye
\*\*
\centerline{\titolo References}

\*
\halign{\hbox to 1.2truecm {[#]\hss} &
        \vtop{\advance\hsize by -1.25 truecm \0#}\cr

BG1& {A. Berretti, G. Gentile:
Scaling properties for the radius of convergence of a
Lindstedt series: the standard map,
{\it J. Math. Pures Appl.}, to appear. }\cr
BG2& {A. Berretti, G. Gentile:
Bryuno function and the standard map, preprint. }\cr
CGJ& {C. Chandre, M. Govin, H.R. Jauslin:
Kolmogorov-Arnold-Moser renormalization-group approach to
the breakup of invariant tori in Hamiltonian systems,
{\it Phys. Rev. E} {\bf 57} (1998), (2), 1536--1543. }\cr
CGJK& {C. Chandre, M. Govin, H.R. Jauslin, H. Koch:
Universality of the breakup of invariant tori in Hamiltonian flows,
{\it Phys. Rev. E} {\bf 57} (1998), (6), 6612--6617. }\cr
C& {V. Chirikov:
A universal instability of many-dimensional oscillator systems,
{\it Phys. Rep.} {\bf 52} (1979), 263--379. }\cr
D1& {A.M. Davie:
The critical function for the semistandard map,
{\it Nonlinearity} {\bf  7} (1994), 219--229. }\cr
D2& {A.M. Davie:
Renormalization for analytic area-preserving maps, unpublished. }\cr
ED& {D.F Escande, F. Doveil:
Renormalization method for computing the threshold of the large-scale
stochastic instability in two degrees of freedom Hamiltonian systems,
{\it J. Stat. Phys.} {\bf 26} (1981), 257--284. }\cr
GGM& {G. Gallavotti, G. Gentile, V. Mastropietro:
Field Theory and KAM tori,
{\it Math. Phys. Electron. J.} {\bf 1} (1995), paper 5, 1--13. }\cr
GM& {G. Gentile, V. Mastropietro:
Tree expansion and multiscale decomposition for KAM tori,
{\it Nonlinearity} {\bf 8} (1995), 1--20. }\cr
M1& {R.S. MacKay:
A renormalization approach to invariant circles in area-preserving maps,
{\it Physica D} {\bf  7} (1983), 283--300. }\cr
M2& {R.S. MacKay:
{\sl Renormalization in area-preserving maps},
World Scientific, London (1993). }\cr
Y& {J.-C. Yoccoz:
Th\'eoreme de Siegel, nombres de Brjuno and p\^olinomes quadratiques,
{\it Ast\'erisque} {\bf 231}, 3--88 (1995). }\cr
}

\bye